\documentclass[preprint,12pt]{elsarticle}

\usepackage{graphics}
\usepackage{epsfig}

\usepackage{amssymb}

\journal{Physics Letters B}

\begin{document}

\begin{frontmatter}



\title{Metamaterials Mimicking Dynamic Spacetime, D-brane and
Noncommutativity in String Theory}


\author{Rong-Xin Miao,\ Rui Zheng,\ Miao Li}

\address{Kavli Institute for Theoretical Physics, Key Laboratory
of Frontiers in Theoretical Physics, Institute of Theoretical
Physics, Chinese Academy of Sciences, Beijing 100190, People's
Republic of China, Interdisciplinary Center for Theoretical Study,
University of Science and Technology of China, Hefei, Anhui 230026,
China}

\begin{abstract}
We propose a scheme to mimic the expanding cosmos in $1+2$
dimensions in laboratory. Furthermore, we develop a general
procedure to use nonlinear metamaterials to mimic D-brane and
noncommutativity in string theory.
\end{abstract}


\end{frontmatter}


\section{Introduction}
String theory is a developing field of modern theoretical physics
which attempts to unify quantum mechanics and general relativity.
The basic objects in string theory are one dimensional strings and
high dimensional D-branes. To be self-consistent, the theory also
requires supersymmetry, extra dimensions. The effective geometry on
a D-brane becomes noncommutative when there is a nonzero background
B field \cite{Witten}. String theory has made tremendous theoretical
progresses in the past 15 years, such as  explaining microscopic
origin of the black hole entropy \cite{Strominger} in certain cases,
the discovery of AdS/CFT \cite{Maldacena}. However, string theory
remains hard to be tested by experiments due to the extremely high
energy scale.

It will be of great interest if one can mimic some phenomena of
string theory in laboratory.  Fortunately, the developments of
metamaterials and transformation optics
\cite{exp,Houck,Smith,Cubukcu,Yen,Linden,Leo,Ge,Huangyang1,JohnPendry11,JohnPendry12,all}
may be helpful for us to achieve this goal. Recently, metamaterials
were used to design various interesting devices such as
electromagnetic cloak \cite{Pendry,Leonhardt}, perfect lens
\cite{Pendry1}, illusion devices\cite{Chen11,Chen12} and so on. They
can even be used to make an artificial black hole
\cite{Narimanov,Cheng:2009ja,Huanyang} and to mimic cosmos
\cite{Miao Li1,Miao Li2}. For example, in recent works \cite{Miao
Li1,Miao Li2}, we find that the Casimir energy of the
electromagnetic field in de Sitter space is proportional to the size
of the horizon, the same form of the holographic dark energy
\cite{Li}. We suggest to make metamaterials to mimic static de
Sitter space in laboratory and measure the predicted Casimir energy.

In this paper we move on and propose to use metameterials to mimic
some phenomena in cosmology and string theory. We design an approach
to mimic a dynamic space-time in $1+2$ dimensions by materials with
constant permeability and varying permittivity, to avoid the complex
issue of designing metameterials with synchronously varying
permittivity and permeability \cite{Ginis}. Our proposal can be
implemented by 3-level atoms or a nonlinear dielectric. Besides, we
further suggest to use nonlinear metamaterials to mimic D-brane and
noncommutative geometry in string theory. This topic is
scientifically interesting, since it implies the possibility of
testing and investigating some aspects of string theory at low
energy scale in laboratory. Besides, the various properties of
D-brane, noncommutativity and nonlinear electrodynamics will
motivate us to design interesting optical devices.

This paper is organized as follows. In Sec. II, we develop an
executable experimental procedure to mimic expanding cosmos in $1+2$
dimensions with abundant interesting effects such as redshift, the
cosmic microwave background (CMB) and the future event horizon. In
Sec. III, we develop in theory a general procedure to use
metamaterials to mimic gravity, D-brane, noncommutativity and so on.
We conclude in Sec. IV.

\section{Mimic dynamic spacetime in 1+2 dimension}
In this section we discuss the issue of mimicking a $1+2$
dimensional dynamic space-time in metameterials. To do this let us
consider the analogy between curved spacetime and dielectric medium
for electromagnetic fields, as was explicitly studied in
\cite{Leo,Pendry}. Especially, it is proven that light has the same
behavior in dielectric medium and curved spacetime, under the
conditions that the permittivity and the permeability are related to
the metric by
\begin{equation}
     \varepsilon^{ij}=\frac{\varepsilon_0\sqrt{-g}}{-g_{00}}g^{ij}, \  \mu^{ij}=\frac{\sqrt{-g}}{-\varepsilon_0 c^2g_{00}}g^{ij}\ .\label{perpea}
\end{equation}
It should be stressed that this equation is valid only in 1+3
dimensions. This relation can be used to design metameterials and
mimic specific kind of spacetime. For example, to mimic an expanding
universe, the authors of \cite{Ginis} proposed to vary the
permeability by split-ring resonators, and change the permittivity
by electrooptical modulation.

Theoretically, one can make use of Eq.(\ref{perpea}) to mimic an
arbitrary spacetime. However, to mimic a dynamic spacetime, for
instance, an expanding universe, one should simultaneously control
the permeability and the permittivity to ensure them varying
synchronously. Unfortunately, this is a hard task for experimental
physics. To avoid this problem in this paper we just consider the
case of mimicking a 1+2 dimensional spacetime. In this section we
study two models, including a real 1+2 dimensional spacetime and a
1+2 dimensional subspace of the 1+3 dimensional spacetime. In both
cases we find that the permittivity and the permeability can be
independent of each other, and the above problem is automatically
resolved. Besides, it is natural and useful to consider the
simplified case of realizing the 1+2 dimensional spacetime before
turning to the much more complicated case of realizing a 1+3
dimensional spacetime. The metric of a homogeneous and isotropic
universe in 1+2 dimensions takes the form
\begin{equation}\label{metric}
ds^{2}=g_{00}c^2dt^{2}+g_{11}dx^{2}+g_{22}dy^{2}\ ,
\end{equation}
with $g_{11}=g_{22}$. Using the method developed in the next section
we find that the dielectric with the following permeability and
permittivity can be used to mimic this 1+2 dimensional metric

\begin{equation}\label{2}
\mu=\frac{\sqrt{-g}}{-\varepsilon_0c^2g_{00}},\ \ \
\varepsilon^{ij}=\frac{\varepsilon_0\sqrt{-g}}{-g_{00}}g^{ij}\
(i,j=1,2)\ .
\end{equation}

It is clear that the permittivity and the permeability are
independent of each other. This remarkable property can be used to
simplify experimental design. In fact, to mimic the expanding
cosmos, one can keep the permeability as a constant and only vary
the permittivity. We rewrite the metric in terms of the permittivity
and the permeability from Eq.(\ref{2})
\begin{equation}
ds^2=-\frac{\varepsilon_0^2}{\varepsilon^2(t)}c^2dt^2+\frac{\mu(t)\varepsilon_0^2c^2}{\varepsilon(t)}(dx^{2}+dy^{2})\
.
\end{equation}
To realize the Friedmann-Robertson-Walker(FRW) metric
\begin{equation}\label{FRW}
ds^{2}=-c^2d\tau^{2}+a^{2}(\tau)(dx^{2}+dy^{2})\ ,
\end{equation}
we redefine the time coordinate as $d\tau=\varepsilon_0
dt/\varepsilon(t)$. Correspondingly we should express all the
parameters in terms of this new time $\tau$ rather than the
laboratory time $t$. For example, $\mu(\tau)=1/(\varepsilon_0 c^2)$
and $\varepsilon(\tau)=\varepsilon_0/a^{2}(\tau)$.

Note that the Maxwell theory in 1+2 dimensions can be derived from
that in 1+3 dimensions by imposing the self-consistent conditions
$E_{3}=B^{1}=B^{2}=0$ and
$\partial_{3}E_{1}=\partial_{3}E_{2}=\partial_{3} B^{3}=0$. Hence,
we can model this 1+2 dimensional metric in dielectric with
translation invariance along the z-axis and use transverse cross
section perpendicular to the z-axis to mimic the two-dimensional
space we are interested in. One may polarize photons to keep the
magnetic field along z-axis, so that the magnetic field acts as a
scalar on the cross section. Meanwhile, to confine them completely
on the cross section, it is required that photons have no momentum
along the axis. Therefore, one can make an analogy between the
electromagnetic fields on the cross section and the counterparts in
1+2 dimensional spacetime
\begin{equation}
\tilde{E}_{i}\sim E_{i}\mbox{ }(i=1,2),\mbox{ }\mbox{
}\tilde{B}_{3}\sim B \ .
\end{equation}
Accordingly, one can also make an analogy between the relevant
parameters of the dielectric and those in 1+2 dimensions
\begin{equation}
\tilde{\varepsilon}^{ij}\sim \varepsilon^{ij}\mbox{
}(i,j=1,2),\mbox{ }\mbox{ }\tilde{\mu}^{33}\sim \mu \ .
\end{equation}
Thus, to mimic the metric (\ref{FRW}), one only need to design
metameterials with the property
\begin{equation}
\tilde{\varepsilon}^{ij}\mbox{ }=\varepsilon_0a^{-2}(\tau)\
\delta^{ij}\ \ (i,j=1,2),\ \ \tilde{\mu}^{33}=1/(\varepsilon_0 c^2)\
.
\end{equation}

We can also make the electric field parallel to the z-axis so that
the electric field becomes a scalar on the cross section. In this
case we can not mimic the Maxwell theory in 1+2 dimensions in which
electric fields are vectors. Instead, we can use it to mimic the
Maxwell theory in a 1+2 dimensional subspace of the 1+3 dimensional
spacetime. As in the above case, to be confined on the cross
section, photons should have no momentum along the z-axis. When
electric field is polarized along the axis, the nonzero components
of the electromagnetic fields are $E_{3}$ and $B_{ij}\mbox{
}(i,j=1,2)$, and the relevant dielectric parameters are
$\varepsilon^{33}$ and $\mu^{ij}\mbox{ }(i,j=1,2)$. It should be
stressed that if we consider only the 1+2 dimensional subspace, the
irrelevant dielectric parameters such as $\varepsilon^{ij}$ and
$u^{33}$ need not take the exact form (\ref{perpea}). In fact, they
can take arbitrary forms and do not affect the polarized photons on
the cross section. Setting $g_{33}=1$, one can derive the relevant
dielectric parameters from Eq.(\ref{perpea}) as
\begin{equation}
\varepsilon^{33}=\frac{\varepsilon_0\sqrt{-g}}{-g_{00}}\ ,\ \
\mu^{ij}=\frac{\sqrt{-g}}{-\varepsilon_0
c^2g_{00}}g^{ij}\label{para2}\ (i,j=1,2)\ .
\end{equation}
 We have the metric of the 1+2 dimensional
subspace from Eq.(\ref{para2})
\begin{equation}
ds^2=-\frac{1}{\varepsilon_0^2
c^2\mu^2(t)}dt^2+\frac{\varepsilon^{33}(t)}{\varepsilon_0^2
c^2\mu(t)}(dx^{2}+dy^{2})\ .
\end{equation}
Next we set $\mu=1/(\varepsilon_0 c^2)$ and denote
$\varepsilon^{33}(t)$ by $\varepsilon(t)$,
 it follows that
\begin{equation}\label{design2}
ds^2=-c^2dt^2+\frac{\varepsilon(t)}{\varepsilon_0}(dx^{2}+dy^{2})\ .
\end{equation}
where $t$ is just the laboratory time. The scale factor, which
signifies the scale of the universe, is
$a(t)=\sqrt{\varepsilon(t)/\varepsilon_0}$. Now we get the FRW
metric in 1+2 dimensional flat universe. One can mimic an expanding
universe using metameterials with increasing $\varepsilon(t)$.

In the rest of this section, we focus on the experimental design
(\ref{design2}) with the electric field polarized along the axis.
Experimentally one has several alternative methods to vary the
permittivity, e.g., by making use of the electrooptical effect and
the model consisting of $N$ identical 3-level atoms \cite{3-level}
(the second method allows a comparably wider range of variation).
Many interesting effects, such as the future event horizon, redshift
and the varying CMB temperature, are expected in the experiment.

If the permittivity increases fast enough, a future event horizon
will arise
\begin{equation}
r_{h}(t)=a(t)\int_t^\infty\frac{c
dt'}{a(t')}=\sqrt{\varepsilon(t)}\int_t^\infty\frac{c
dt'}{\sqrt{\varepsilon(t')}}\ .
\end{equation}
This is the largest distance that photons can travel. In the case of
inflation, $a(t)=e^{Ht}$ ($H$ is the Hubble constant), and
$r_{h}(t)=c H^{-1}$. Thus $c H^{-1}$ is the limited distance light
rays can propagate (the optical distance, not the coordinate
distance).

For light with frequency $\nu_{0}$ emitted at time $t_0$, its
frequency at time $t_{1}$ takes the form \cite{weinberg}
\begin{equation}
\nu_1/\nu_0=a(t_0)/a(t_1)=\sqrt{\varepsilon(t_0)/\varepsilon(t_1)}\
.
\end{equation}
If one increases $\varepsilon(t)$, the frequency decreases
accordingly and light is redshifted. The number density of photons
is described by the Bose-Einsein distribution
\begin{equation}
n(\nu,t)d\nu=\frac{8\pi\nu^2d\nu/c^3}{\exp(h\nu/k_\mathcal
{B}T(t))-1}\ .
\end{equation}
Integrating this equation, one finds that the total photon number
and the total energy  $N \propto T^3$ and $U\propto T^4$,
respectively. Thus the average energy of photons takes the form
$\bar{E}=h\bar{\nu}=U/N\propto T$. Notice that the temperature
varies as $T\propto \bar{\nu}\propto 1/a(t)\propto
1/\sqrt{\varepsilon(t)}$, which is exactly the same form of the CMB
temperature.

\section{Mimic D-brane and Noncommutativity in String Theory}
In this section we demonstrate the feasibility of mimicking
generalized electrodynamics by metamaterials. Let us start with the
Maxwell equations in a dielectric medium
\begin{equation}\label{Max}
\partial_{i}B^{i}=0,\ \
\epsilon^{ijk}\partial_{j}E_{k}+\partial_{t}B^{i}=0\ ,
\end{equation}
\begin{equation}
\partial_{i}D^{i}=0,\ \
\epsilon^{ijk}\partial_{j}H_{k}-\partial_{t}D^{i}=0 \ .\label{Maxeq}
\end{equation}
 where the electric displacement $D^{i}$ and magnetizing field $H_{i}$
are related to the electric field $E_{i}$ and the magnetic field
$B^{i}$ through
\begin{equation}\label{constitutive}
D^{i}=\varepsilon^{ij}E_{j},\ \ \ B^{i}=\mu^{ij}H_{j}.
\end{equation}
For generalized electrodynamic theory in an arbitrary background
spacetime, the Lagrangian takes the form $L(F_{\mu\nu},g_{\mu\nu})$.
From the definition of the field strength $F_{\mu\nu}=\partial_\mu
A_\nu - \partial_\nu A_\mu$, one can derive the Bianchi identity
\begin{equation}\label{Bianchi}
\partial_\mu F_{\nu\rho}+\partial_\nu F_{\rho\mu}+\partial_\rho
F_{\mu\nu}=0\ .
\end{equation}
By defining
\begin{equation}
E_{i}=F_{i0}\ ,\ \ B^{i}=\frac{1}{2c}\epsilon^{ijk}F_{jk}\ ,
\end{equation}
one can rewrite the Bianchi identity Eq.(\ref{Bianchi}) as
Eq.(\ref{Max}). From the equations of motion (Eq.(\ref{Eq: E-L}) in
the Appendix), one can derive Eq.(\ref{Maxeq}) as long as the
electric displacement and magnetic field are defined by
\begin{equation}\label{DH}
D^{i}=\frac{\partial L(F_{\mu\nu},g_{\mu\nu})}{\partial E_{i}},\\
\mbox{ }\mbox{ }H_{i}=-\frac{\partial
L(F_{\mu\nu},g_{\mu\nu})}{\partial B^{i}}.
\end{equation}
For details of the derivation please refer to the Appendix. Clearly,
in the framework of the generalized electrodynamical theory, photons
have the same behavior as those in the dielectric medium satisfying
the constitutive relations (\ref{DH}), due to the same expression of
Maxwell equations in these two cases. This similarity implies a
promising method to mimic generalized electrodynamics, D-brane and
noncommutativity in string theory.

Following our procedure, the well-known analogy (\ref{perpea})
between curved spacetime and dielectric can be easily derived. For
simplicity, here we take the case of  1+2 dimensions as an example.
The Lagrangian of the electromagnetic field in curved space is
\begin{equation}
L(F)=-\frac{1}{4
}\varepsilon_0\sqrt{-g}F^2=-\frac{1}{4}\varepsilon_0\sqrt{-g}g^{\mu\alpha}g^{\nu\beta}F_{\mu\nu}F_{\alpha\beta}\
,
\end{equation}
from which one can derive
\begin{equation}
D^{i}=\frac{\partial L}{\partial E_{i}}=\frac{\partial L}{\partial
F_{i0}}=\frac{\varepsilon_0\sqrt{-g}}{-g_{00}}g^{ij}E_j\ ,
\end{equation}
\begin{equation}
H=-\frac{\partial L}{\partial B}=c\frac{\partial L}{\partial
F_{21}}=\frac{-\varepsilon_0 c^2g_{00}}{\sqrt{-g}}B\ .
\end{equation}
Comparing the above equations with constitutive relations
(\ref{constitutive}), one gets
\begin{equation}
\varepsilon^{ij}=\frac{\varepsilon_0\sqrt{-g}}{-g_{00}}g^{ij} ,\ \
\mu=\frac{\sqrt{-g}}{-\varepsilon_0 c^2g_{00}}\ .\label{perper12}
\end{equation}

With this procedure at hand, we are in a position to investigate
some interesting examples. In the following paragraphs, we study the
mimicking of single $D_{p}$-brane and noncommutativity in string
theory.

In string theory, the low-energy dynamics of single $D_{p}$-brane is
described by the Born-Infeld action
\begin{equation}
S=-\varepsilon_0T_{D_{p}}\int
d^{p+1}x\sqrt{-\det(\eta_{\mu\nu}+2\pi\alpha'F_{\mu\nu})}\ ,
\end{equation}
where $T_{D_{p}}$ is the tension, $p+1$ is the number of dimensions
of the world-volume. From the above arguments, we can use an
appropriate dielectric to mimic the low-energy dynamics of
$D_{p}$-brane. Take $p=2$ as an example, the corresponding
dielectric has the following permittivity and permeability
\begin{equation}\label{Dbrane}
\varepsilon=\frac{4\varepsilon_0T_{D_{p}}\pi^2\alpha'^2}{\sqrt{1+4\pi^2\alpha'^2(c^2{B}^2-\vec{E}^2)}}=\mu^{-1}/c^2\
.
\end{equation}
As in the case of 1+2 dimensional expanding universe, one can mimic
this $D_{2}$-brane by dielectric with translation invariance along
the z-axis and impose the self-consistent conditions
$E_{3}=B^{1}=B^{2}=0$,
$\partial_{3}E_{1}=\partial_{3}E_{2}=\partial_{3} B^{3}=0$.

We encourage experimentalists to design such nonlinear
metamaterials, they have great applications in mimicking string
theory in laboratory. In such nonlinear metamaterials, a mass of
novel phenomena are expected. For example, there is a upper limit of
electric field intensity, the self-energy of the charge is finite,
there is no birefringence and so on. To keep the permittivity and
the permeability (\ref{Dbrane}) real, it is clear that $E\leq1/(2\pi
\alpha')$. In general, when these parameters become complex, the
system will exhibit an instability with abundant photons production
and absoption. There is a correspondence of such instability in
string theory. As we know, D-branes are objects where open strings
can end, and charges carried by the two ends of a string are
opposite. Thus, for an open string with both ends on the D-brane, an
electric field will pull the two endpoints of string apart. If $E$
is large enough, it will overcome the string tension to tear apart
the string. It implies an instability similar to that in the
dielectric. This instability of D-brane is hard to be observed in
laboratory, but it is possible to mimic this effect by metamterials.
Besides, metamaterials seem to have the unique capability to mimic
such novel phenomenons.

In string theory the effective geometry becomes noncommutative (NC)
when there is a background $B$-field \cite{Witten}
\begin{equation}\label{noncommutative}
[x^{\mu}, x^{\nu}]=i\mbox{ } \theta^{\mu\nu}\ ,
\end{equation}
where $x^{\mu}$ are spacetime coordinates and $\theta_{\mu\nu}$ is a
constant antisymmetric tensor. The generalized Yang-Mills action in
NC space is
\begin{equation}\label{YM}
S=-\frac{1}{4}\varepsilon_0\int d^{4}x \mbox{
}\hat{F}_{\mu\nu}\ast\hat{F}^{\mu\nu}\ ,
\end{equation}
where the product $*$ and field strength $F_{\mu\nu}$ are defined as
\begin{equation}\label{product}
f(x)\ast
g(x)=exp(\frac{i}{2}\theta^{\mu\nu}\frac{\partial}{\partial\xi^{\mu}}\frac{\partial}{\partial\zeta^{\nu}})f(x+\xi)g(x+\zeta)|_{\xi=\zeta=0}\
,
\end{equation}
\begin{equation}\label{F}
\hat{F}_{\mu\nu}=\partial_{\mu}\hat{A}_{\nu}-\partial_{\nu}\hat{A}_{\mu}-i
[\hat{A}_{\mu},\hat{A}_{\nu}]_{\ast}\ .
\end{equation}

In general, the NC Yang-Mill theory can be converted into a
commutative but nonlinear counterpart by the Seiberg-Witten map. For
example, when $\hat{F}_{\mu\nu}$ is a constant, the Seiberg-Witten
map is
\begin{equation}\label{map}
\hat{F}_{\mu\nu}=\frac{F_{\mu\nu}}{1+\sqrt{\frac{\varepsilon_0}{\hbar
c}}F^{\alpha\beta}\theta_{\alpha\beta}}\mbox{ }\ ,
\end{equation}
where $F_{\alpha\beta}$ is the ordinary field strength in
commutative space.  From the procedure developed in this section,
one can mimic the NC Maxwell theory by metamaterials satisfying

\begin{eqnarray}\label{NC}
D^i&=&\frac{\varepsilon_0E^i}{(1+2\sqrt{\frac{\varepsilon_0}{\hbar
c}}E_l\theta^{l0}+\epsilon_{lmn}\sqrt{\frac{c\varepsilon_0}{\hbar
}}\theta^{lm}B^n)^2}\nonumber\\
&+&\frac{2\sqrt{\frac{\varepsilon_0}{\hbar
c}}\varepsilon_0(c^2\vec{B}^2-\vec{E}^2)\theta^{i0}}{(1+2\sqrt{\frac{\varepsilon_0}{\hbar
c}}E_l\theta^{l0}+\epsilon_{lmn}\sqrt{\frac{c\varepsilon_0}{\hbar
}}\theta^{lm}B^n)^3}\ ,
\nonumber\\
H_i&=&\frac{\varepsilon_0
c^2B_i}{(1+2\sqrt{\frac{\varepsilon_0}{\hbar
c}}E_l\theta^{l0}+\epsilon_{lmn}\sqrt{\frac{c\varepsilon_0}{\hbar
}}\theta^{lm}B^n)^2}\nonumber\\
&-&\varepsilon_0
c\frac{(c^2\vec{B}^2-\vec{E}^2)\epsilon_{ijk}\sqrt{\frac{\varepsilon_0}{\hbar
c}}\theta^{jk}}{(1+2\sqrt{\frac{\varepsilon_0}{\hbar
c}}E_l\theta^{l0}+\epsilon_{lmn}\sqrt{\frac{c\varepsilon_0}{\hbar
}}\theta^{lm}B^n)^3}\ .
\end{eqnarray}

In NC geometry, Lorentz symmetry is broken. This can lead to
birefringence and direction dependent shift in the speed of light.
For simplicity, we consider a constant electromagnetic field with
small fluctuations, thus the Seiberg-Witten map (\ref{map}) for
constant field strength is valid. Replace $F_{\mu\nu}$ with
$f_{\mu\nu}+F_{\mu\nu}$, where $f_{\mu\nu}$ denotes the constant
background field and $F_{\mu\nu}$ is the small fluctuation. The
resulting Lagrangian reduces to
\begin{equation}
L=-\frac{\varepsilon_0}{4(1+\sqrt{\frac{\varepsilon_0}{\hbar
c}}\theta^{\mu\nu}f_{\mu\nu})^2}F_{\mu\nu}F^{\mu\nu}-\frac{\varepsilon_0}{4(1+\sqrt{\frac{\varepsilon_0}{\hbar
c}}\theta^{\mu\nu}f_{\mu\nu})^2}k_{\kappa\lambda\mu\nu}F^{\kappa\lambda}F^{\mu\nu}\
,
\end{equation}
where
\begin{equation}\label{k1}
k_{\alpha\beta\gamma\delta}=\frac{3\frac{\varepsilon_0}{\hbar
c}f_{\mu\nu}f^{\mu\nu}\theta_{\alpha\beta}\theta_{\gamma\delta}}{(1+\sqrt{\frac{\varepsilon_0}{\hbar
c}}\theta^{\mu\nu}f_{\mu\nu})^2}-\frac{2\sqrt{\frac{\varepsilon_0}{\hbar
c}}(f_{\alpha\beta}\theta_{\gamma\delta}+\theta_{\alpha\beta}f_{\gamma\delta})}{1+\sqrt{\frac{\varepsilon_0}{\hbar
c}}\theta^{\mu\nu}f_{\mu\nu}}\ ,
\end{equation}
we keep only terms up to the second order in $F_{\mu\nu}$. The
linear terms are absent due to the equations of motion of the
background field. The equation of motion of the fluctuation is
\begin{equation}
\partial^{\nu}F_{\mu\nu}+k_{\mu\nu\rho\sigma}\partial^{\nu}F^{\rho\sigma}=0\
.
\end{equation}
For a plane wave
$F_{\mu\nu}(x)=F_{\mu\nu}(p)e^{-ip_{\alpha}x^\alpha/\hbar}$, from
the above equation one can get
\begin{equation}
M^{ij}E_{j}=(-\delta^{ij}p^2-p^{j}p^k-2k^{i{\mu\nu}j}p_{\mu}p_\nu)E_{j}=0\
.
\end{equation}
Following the general method, by requiring the determinant of
$M^{ij}$ vanishes, one can derive the dispersion relation of
photons, then obtain the group velocity \cite{LVinED}
\begin{equation}\label{velocity}
\upsilon_g=c|\nabla_{\vec{p}} p^0|=c(1+\rho\pm\sigma)\ .
\end{equation}
 to leading order in $k_{\mu\nu\rho\sigma}$, where
\begin{equation}
\rho=-\frac{1}{2}\tilde{k}^\alpha_\alpha,\ \
\sigma^2=\frac{1}{2}(\tilde{k}_{\alpha\beta})^2-\rho^2\ ,
\end{equation}
with
\begin{equation}\label{k2}
\tilde{k}^{\alpha\beta}=k^{\alpha\mu\beta\nu}\hat{p}_\mu\hat{p}_\nu,\
\ \hat{p}^\mu=p^\mu/|\vec{p}|\ .
\end{equation}

As mentioned above, the velocity of light is direction dependent.
Furthermore, since two solutions exist for $\vec{E}$, it implies the
birefringence effect, where these two components of light propagate
independently with different velocities.

To end this section, we list the main results in the leading order
in the NC parameters and study an interesting optical phenomenon in
such a NC space. It is likely that photons will spend less time
traveling along polyline than along straight line. For simplicity,
we focus on the case that all NC parameters vanish, other than

$\theta^{03}=\theta>0$. To leading order of $\theta$,
 the constitutive relations (\ref{NC}) of metamaterials become,
\begin{eqnarray}
H_i&=&\varepsilon_0c^2(1+4\sqrt{\frac{\varepsilon_0}{\hbar c}}\theta
E_3)B_i,\nonumber\\
D^i&=&\varepsilon_0(1+4\sqrt{\frac{\varepsilon_0}{\hbar c}}\theta
E_3)E^i-2\varepsilon_0(c^2\vec{B}^2-\vec{E}^2)\sqrt{\frac{\varepsilon_0}{\hbar
c }}\theta\delta^{i}_{3}\ .
\end{eqnarray}
For simplicity, we consider only the case $\vec{E}=0$,
$\vec{B}=(0,0,B)$\ for the background field. Focus on the first
order velocity (\ref{velocity}), from Eq.(\ref{k1}) and
Eq.(\ref{k2}) we learn that only $\hat{p}^\mu$ in the leading order
of $\theta$ contribute, thus we can take
$\hat{p}^\mu=(1,\sin\varphi,0,\cos\varphi)$ in x-z plane as a good
approximation $(\varphi\in[0,\pi/2]$ is the angle between the
momentum of photons and background magnetic field). The velocity of
photons then becomes,
\begin{equation}\label{group}
\upsilon_g=c\pm2c\sqrt{\frac{c\varepsilon_0}{\hbar }}\theta
B\sqrt{(1+\cos^{2}\varphi)\sin^{2}\varphi}\ .
\end{equation}
We focus on the case
$\upsilon_g=c+2c\sqrt{\frac{c\varepsilon_0}{\hbar }}\theta
B\sqrt{(1+\cos^{2}\varphi)\sin^{2}\varphi}$ below. It is clear that
this velocity increases monotonically with $\varphi\in[0,\pi/2]$,
thus the velocity assumes its minimum value $\upsilon_g=c$ when
photons propagate along the direction of the magnetic field. It
should be stressed that though the group velocity (\ref{group})
generally exceeds the speed of light $c$ in vacuum, it does not mean
that information can transfer faster than $c$. Various experiments
have verified that it is possible for the group velocity of light in
materials significantly exceed the speed of light in vacuum.
However, in all these cases, the speed of information remains less
than or equal to $c$ \cite{speed}.
\begin{figure}
\includegraphics[width=6cm]{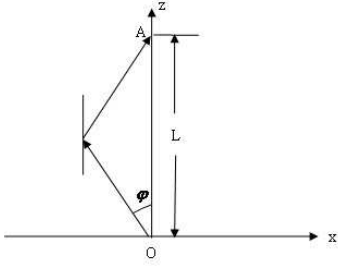}
\caption{The polyline and straight line between $O$ and $A$.}
\end{figure}

As depicted in Fig.1, the distance between these two points $O$ and
$A$ is $L$. The time that photons will spend to cover this distance
along the straight line is
\begin{equation}
t_0=L/c\ .
\end{equation}
Photons can also travel along another path, that is, the polyline
depicted in Fig.1. In this case, photons are reflected to point $A$
by a mirror, and the time it will take to travel along this polyline
is
\begin{equation}
t_1=\frac{L/\cos\varphi}{c+2c\sqrt{\frac{c\varepsilon_0}{\hbar
}}\theta B\sqrt{(1+\cos^{2}\varphi)\sin^{2}\varphi}}\ .
\end{equation}
One can always find a nonempty interval for $\varphi$ to make sure
$t_1<t_0$. For instance, set $\sqrt{\frac{c\varepsilon_0}{\hbar
}}\theta B=0. 1$, one can find $c t_1/L$ varies with $\varphi$ as
depicted in Fig.2.
\begin{figure}
\includegraphics[width=6cm]{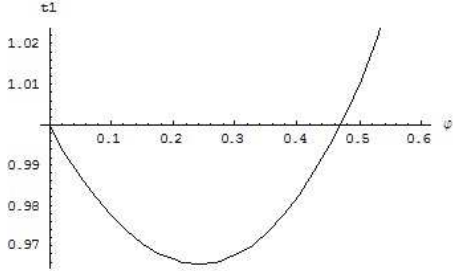}
\caption{The time photons spend along the polyline with respect to
the direction of propagation.}
\end{figure}
Hence, it takes less time for photons to move along the polyline
than that along the straight line for $\varphi\in(0,0.47)$. And the
fastest path
 for photons to travel to point $A$ is along the angle
 $\varphi=0.24$. Thus, the least time for photons to travel from point $O$ to point $A$
 is $t_1=0.97L/c$.

It should be stressed that in such a NC space with a constant
background magnetic field, free photons still travel along straight
line, but the fastest path for light to propagate between two points
is generally a polyline. This interesting, counter-intuitive effect
is due to Lorentz violation, which is believed to serve as signals
of new physics coming from the Planck scale.  Since the direct
detection of Lorentz violation is difficult for the time being, it
is of great significance if one can mimic and detect this effect in
laboratory.

\section{Conclusion}
In this paper we discuss the issue of using metameterials to mimic
some phenomena in cosmology and string theory. We propose a simple
procedure to mimic a $1+2$ dimensional expanding universe. We find
that this can be implemented by materials with constant permeability
and varying permittivity, so that the troublesome problem of
designing metameterials with synchronously varying permittivity and
permeability is automatically avoided. Moreover, we further suggest
to use nonlinear metamaterials to mimic D-brane and noncommutative
geometry in string theory. The permittivity and permeability of the
corresponding metamaterials are calculated and provided in Sec. III.
We see that it is possible to test and investigate some aspects of
string theory at low energy scale in laboratory. However, it should
be stressed that our proposals are as yet in theory, the
metamaterials which we suggest to mimic D-brane and noncommutative
geometry are extremely difficult to make. We hope that with the
progress in the field of metamaterials our proposals will be
realized in the future.

\section*{Acknowledgments}
We would like to thank Prof. Mo Lin Ge for informing us of the
exciting developments in the field of electromagnetic cloaking and
metamaterials. We also thank Tao Shi and Xiao-Dong Li for valuable
discussions and suggestions. This work was supported by the NSFC
grant No.10535060/A050207, a NSFC group grant No.10821504 and
Ministry of Science and Technology 973 program under grant
No.2007CB815401.

\section*{Appendix}
In this appendix, we show that with definition (\ref{DH}), the
Maxwell equations in generalized electrodynamic theory and
dielectric coincide.

For a generalized electrodynamic theory, the Euler-Lagrange equation
takes the form
\begin{equation}
\label{Eq: E-L}
\partial_{\mu}\frac{\partial
L}{\partial (\partial_{\mu}A_{\nu})}-\frac{\partial L}{\partial
A_{\nu}}=0\ .
\end{equation}
Note that
\begin{equation}
D^{i}=\frac{\partial L}{\partial E_{i}}=\frac{\partial L}{\partial
(\partial_{i}A_{0})}\ ,\ \ \ \ \  \frac{\partial L}{\partial
A_{0}}=\frac{\partial L}{\partial(\partial_{0} A_{0})}=0\ ,
\end{equation}
one can easily obtain
\begin{equation}\label{MEQ1}
\partial_{i}D^{i}=\partial_{i}\frac{\partial
L}{\partial (\partial_{i}A_{0})}=0\ .
\end{equation}
Similarly, note that
\begin{equation}
H_{k}=-\frac{\partial L}{\partial
B^{k}}=-\frac{1}{2}c\epsilon_{ijk}\frac{\partial L}{\partial
F_{ij}}\ ,
\end{equation}
\begin{equation}
\epsilon^{mnk}H_{k}=-\frac{c}{2}\epsilon^{mnk}\epsilon_{ijk}\frac{\partial
L}{\partial F_{ij}}=-c\frac{\partial L}{\partial
F_{mn}}=\frac{c}{2}\frac{\partial L}{\partial (\partial_{n}A_{m})}\
,
\end{equation}
from the Euler-Lagrange equation, one can derive
\begin{equation}\label{MEQ2}
\epsilon^{mnk}\partial_{n}H_{k}=\frac{c}{2}\partial_{n}\frac{\partial
L}{\partial
\partial_{n}A_{m}}=-\frac{c}{2}\partial_{0}\frac{\partial
L}{\partial \partial_{0}A_{m}}=c\partial_{0}\frac{\partial
L}{\partial E_{m}}=\partial_{t}D^{m}\ .
\end{equation}
Finally, with Eq.(\ref{MEQ1}) and Eq.(\ref{MEQ2}), we get the same
Maxwell equations (\ref{Maxeq}) as in dielectrics. It should be
mentioned that a similar transformation-optical equivalence
relations has been derived in \cite{Ulf1}.









\end{document}